\documentclass[12pt]{iopart}
\pdfoutput=1
\usepackage{graphicx}
\usepackage[latin1]{inputenc}
\usepackage{amsfonts}
\usepackage{amssymb}
\usepackage{color}
\graphicspath{{./images/}}

\begin{document}
\title[A Van-Der-Waals picture for metabolic networks]{A Van-Der-Waals picture for metabolic networks from MaxEnt modeling: inherent bistability \&  elusive coexistence.}
\author{D.De Martino$^{1}$}
\address{$^1$ Institute of Science and Technology Austria (IST Austria), Am Campus 1, Klosterneuburg A-3400, Austria\\}
\begin{abstract}
% In this article  it is shown 
In this work maximum entropy distributions in the space of steady states of metabolic networks are defined upon  constraining the first and second moment of the growth rate.  Inherent bistability of fast and slow phenotypes, akin to a Van-Der Waals picture, emerges  upon considering  control on the average growth (optimization/repression) and its fluctuations (heterogeneity). 
This  is applied to the carbon catabolic core of {\it E coli} where it agrees with some stylized facts on the persisters phenotype and  it provides a quantitative map with metabolic fluxes, opening for the possibility to detect coexistence  from flux data.  Preliminary analysis  on data for {\it E Coli} cultures in standard  conditions shows, on the other hand, degeneracy for the inferred parameters that extend in the coexistence region. 
\end{abstract}

\section{Introduction}
Quantitative modeling cell  metabolism would give essential insights 
on reconstructing cell physiology from molecular mechanisms with wide applications, a quest that shall be addressed within a systemic perspective \cite{kacser1973control}. 
Upon modeling cell growth, an approach based on stochiometric constraints under a steady-state hypothesis for the underlying chemical network, complemented by optimality principles, it works in semi-quantitative way at population level in lab conditions \cite{majewski1990simple},   
it is straightforward  to formulate \cite{kantorovich1940new} and computationally fast to implement \cite{dantzig2016linear}. 
However, several phenomena in biology would require to assess single-cell heterogeneity \cite{altschuler2010cellular} and thus an extension of modeling beyond the population level.  
Nowadays, recent advancements in microfluidic device can give access to growth rate physiology at single cell level \cite{wang2010robust}. In particular measures on isogenic populations in identical conditions give rise to growth rate distributions with unavoidable fluctuations \cite{kennard2016individuality}, that shall reflect metabolic variability \cite{kiviet2014stochasticity}.
Both the issue of a systemic perspective and the one of addressing heterogeneity could get interesting insights from statistical mechanics methods. Recent applications of  Monte Carlo Markov chain algorithms to integral calculus \cite{Turcin:1971} give rise to the possibility of mapping growth rate distributions into an underlying distribution for the metabolic phenotypes, in the most simple way upon recurring to maximum entropy distributions at fixed average growth rate in the space of metabolic steady states \cite{de2016growth}. These in turns provide  quantitative predictions for experimental estimates of metabolic fluxes as well as  for their scaling, correlations and fluctuations \cite{de2017statistical}. Still this variability does not amount to a substantial phenotypic heterogeneity, since a linear constraint on the average growth leads to simple unimodal distributions. 
On the other hand, motivated by the recent threat of antibiotic resistance development \cite{d2011antibiotic}, there is a substantial interest on analzying truly heterogeneous coexistence of persistent phenotypes \cite{   balaban2004bacterial}, whose modeling could get interesting insights
from physics \cite{allen2016antibiotic}.
In general, a small fraction of bacteria  is usually observed to survive  an antibiotic treatment, after which they are able to regrown and restablish an antibiotic-sensitive population, thus indicating that the origin of the resistance of this ``persistent'' small subpopulation does not rely on genotype mutations. 
Persister phenotypes are thus known to coexist in cultures in stressed conditions and show an active, yet no (or very slow) growing, metabolism \cite{kotte2014phenotypic}. 
It has been quantitatively shown that the fitness landscape of {\it E Coli} endows inherent bistability  of the growth rate \cite{deris2013innate} and  it has been  observed that  the central carbon core metabolism shall thus endow a bistable phenotypic landscape where slow and fast growing cells sit on wells separated by a watersheed barrier whose height and positions are ruled by two phenomenological parameters related to the growth rate and the metabolic flux\cite{radzikowski2016bacterial}.  
In this work a simple phenomenological approach is proposed to model quantitatively bistability and coexistence of fast and slow growing phenotypes for the central catabolic core of {\it E. Coli} in terms of maximum entropy distributions. It will be shown that bimodal distributions are retrievec upon constraining the first two moments of the growth rate in the metabolic space, with the two lagrange multipliers in the Boltzmann-Gibbs distribution that play  the role of the control parameters that shape the bistable phenotypic landscape akin to a Van-Der-Waals picture. It will be shown that, despite its simplicity, this framework captures stylized facts of the metabolic persistent phenotype, like an active metabolism and increased rate of respiration and carbon dioxide production (per unit of carbon intake) with respect to the fast growing phenotype. This framework would ideally provide a mean to infer elusive coexistence of persisters from flux data and a preliminary analysis from {\it E. Coli} cultures in standard conditions will be presented, alongside with conclusions.

\section{Results}
\subsection{The Model}
A standard approach to model metabolism during cell growth is to consider the underlying chemical network in a well-mixed steady state in the continuum limit, that is approximately valid for timescales slower than diffusion and metabolic turnover. 
The reaction network is encoded in a matrix of stoichiometric coefficients $S_{i \mu}$ (metabolite $\mu$ in reaction $i$), that linearly relates the time derivative of concentrations $c_\mu$ to enzymatic fluxes $\nu_i$ (mass balance)
\begin{equation}
\dot{c}_\mu = \sum_i S_{i \mu} \nu_i
\end{equation}
Upon considering steady states and taking into account reversibility, kinetic limits and medium capacity in terms of flux bounds, a space of feasible states is defined that is geometrically a convex polytope $P$
\begin{eqnarray}
\sum_i S_{i \mu} \nu_i =0 \\
\nu_i \in [\nu_i^{min},\nu_i^{max} ].
\end{eqnarray}

In order to model cell growth,  a phenomenological biomass production reaction is added in models, whose flux will be denoted  $\lambda$ and whose maximum in given conditions can be evaluated with linear programming.

Here we will consider Maximum Entropy distributions in the space of steady state of a metabolic network upon fixing the first two moments of the growth rate,
$\langle \lambda \rangle$ and $\langle \lambda^2 \rangle$. By a standard variational approach \cite{jaynes1957information}, constraints on the moments are enforced by  Lagrange multipliers ($\beta,\gamma$) as control parameters in Boltzmann-Gibbs distributions 
\begin{equation}
P({\bf \nu}) \propto e^{\beta \lambda({\bf \nu}) + \gamma \lambda^2({\bf \nu})} \qquad \textrm{if}  \quad {\bf \nu }\in P
\end{equation} 
whose analysis is developed in the next section.
\subsection{A Van-Der-Waals picture}

The marginal distribution $q(\lambda)$ of the growth rate from an uniform sampling of the network under exam, the catabolic core of E.coli, has in a good approximation the simplex-like form 
\begin{equation}
q(\lambda) \propto (\lambda_{max}-\lambda)^a
\end{equation}
where $\lambda_{max}$ is the maximum achievable  growth rate, given the conditions (obtainable with linear programming) and $a=D-d-1$, where $D$ is the dimension of the space and $d$ the dimension of the subspace where $\lambda$ (eg $d=0$ is a vertex). Upon considering the aforementioned constraints on the moments, we have for the marginal distribution of the growth rate $p(\lambda) \propto e^{F_{\beta,\gamma}(\lambda)}$, where the rate function (from now on we assume $\lambda_{max}=1$ for simplicity of notations) has the form
\begin{equation}
F_{\beta,\gamma}(\lambda) = \beta \lambda +\gamma \lambda^2 +  a \log (1-\lambda) \quad \lambda\in[0 ,1]
\end{equation}   
The extremal points of this function can be analyzed as a function of the control parameters in simple terms.
There is a maximum in $\lambda=0$ (where $F(0)=0$) iff $F'(0)=\beta-a<0$.
There are other two extremal points where $F'(\lambda_\pm)=0$, i.e. 
\begin{equation}
\lambda_\pm =\frac{2\gamma-\beta \pm \sqrt{(2\gamma+\beta)^2-8\gamma a}}{4\gamma}
\end{equation}
that are real as soon as $\beta > \sqrt{8\gamma a} -2\gamma$, and positive if $\gamma>\gamma_c=a/2$, with $\lambda_+$ maximum and $\lambda_-$ minimum. We have qualitatively a picture similar to a Van-Der-Waals fluid, with the two ``spinodal'' lines ($\gamma>\gamma_c=a/2$)
\begin{eqnarray}
\beta = a \\
\beta = \sqrt{8\gamma a} -2\gamma \\ \nonumber
\end{eqnarray}
that mark the onset of probability maxima that refer to respectively, slow  (below (5)) and fast (above (6)) growing phenotypes. The curves (5,6) thus define a coexistence region with an equilibrium line  given implicitly  by the equation $F(0)=0=F(\lambda_+)$, where the maxima have equal heights, that marks discontinuous transitions and hysteresis, shrinking at the  point ($\gamma_c=a/2,\beta_c=a$). This picture is summarized in the diagram in Fig. 1.
\begin{figure}[h]
\centering
\includegraphics*[width=.7\textwidth,angle=0]{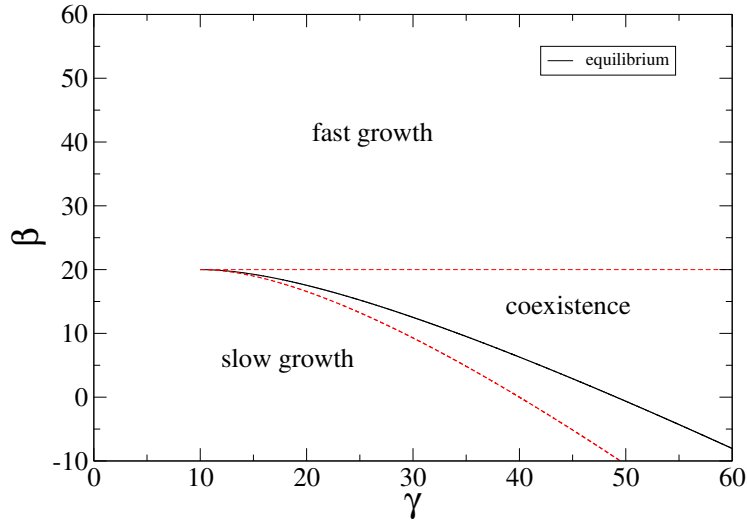}
\caption{Phase diagram of metabolic network states in the plane $(\gamma,\beta)$ of the two lagrange multipliers constraining the first two moments of the growth rate ($a=20$). Red dashed lines: ``spinodal'' lines that mark the onset of probability peaks for slow and fast growing phenotypes. Blak full line: ``equilibrium'' curve where the frequencies of fast and slow phenotypes are equal.}
\label{fig:fig1}
\end{figure}
Growth rate marginal distributions in particular can be obtained analytically by  Legendre transform as we show in fig 2, where  a comparison with Monte Carlo computations on the network under exam (see materials and methods section ) is shown.
\begin{figure}[h]
\centering
\includegraphics*[width=.45\textwidth,angle=0]{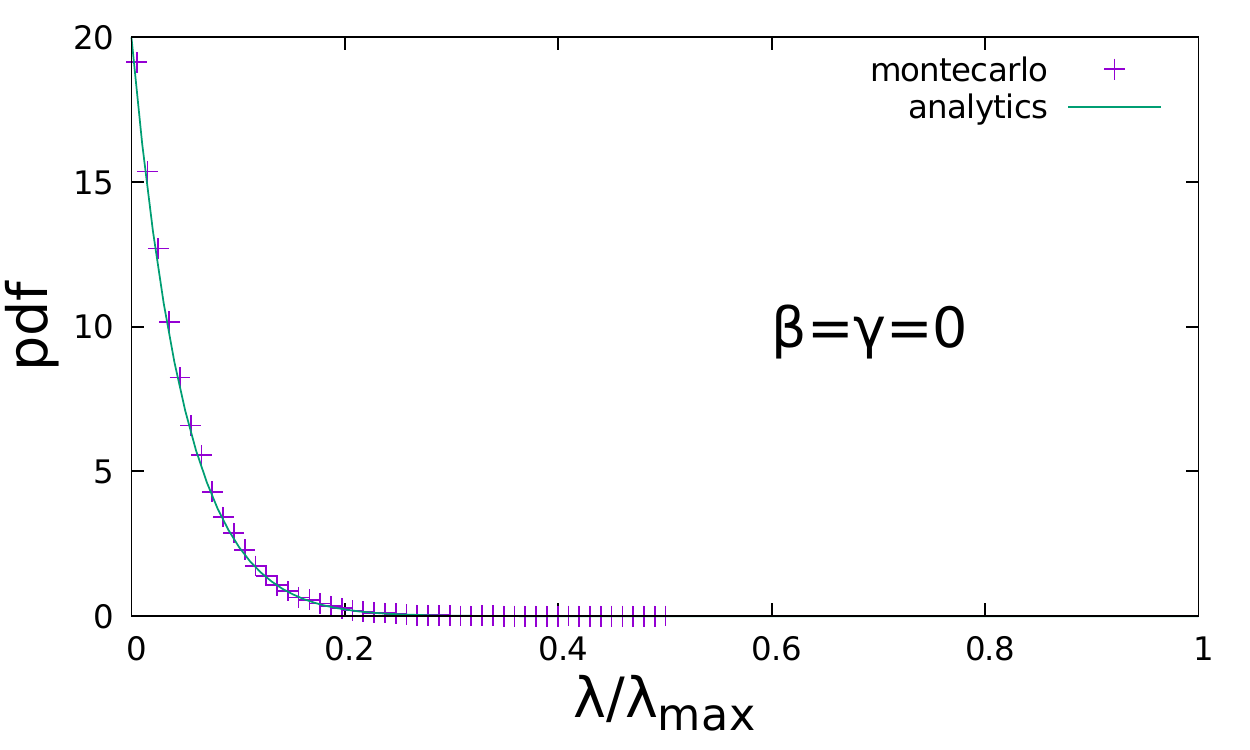}
\includegraphics*[width=.45\textwidth,angle=0]{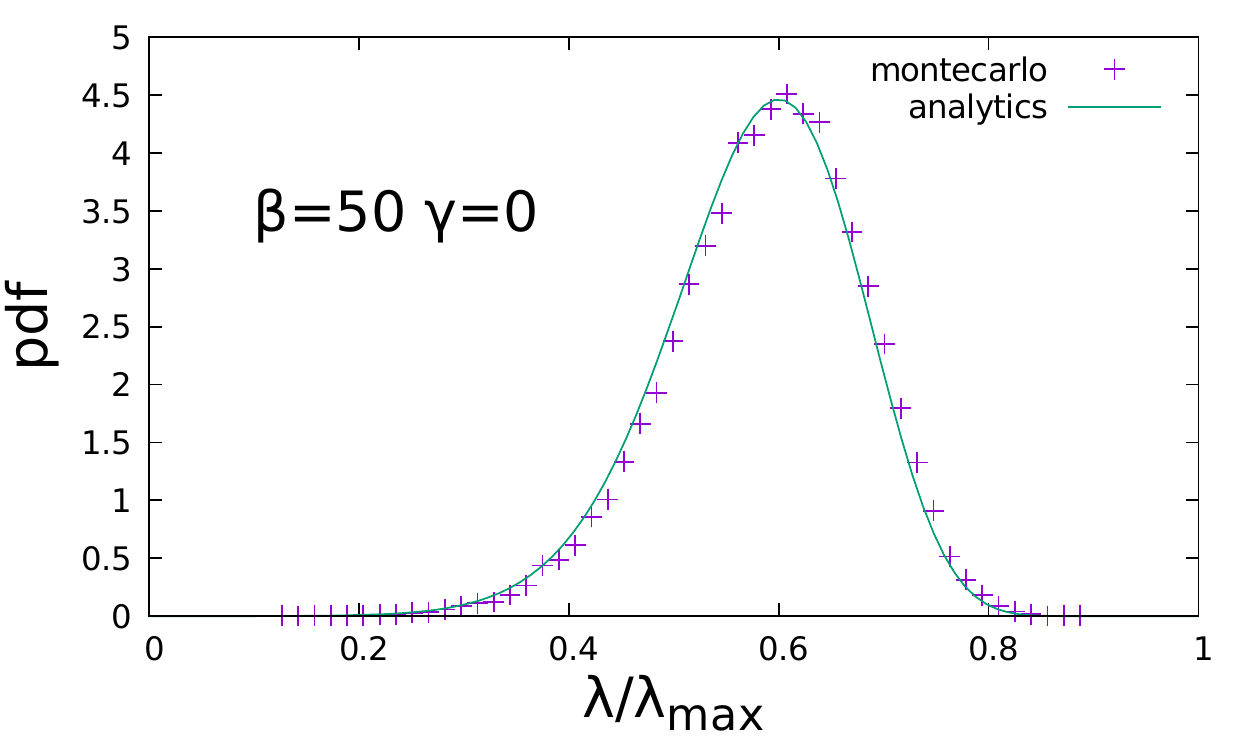}
\includegraphics*[width=.45\textwidth,angle=0]{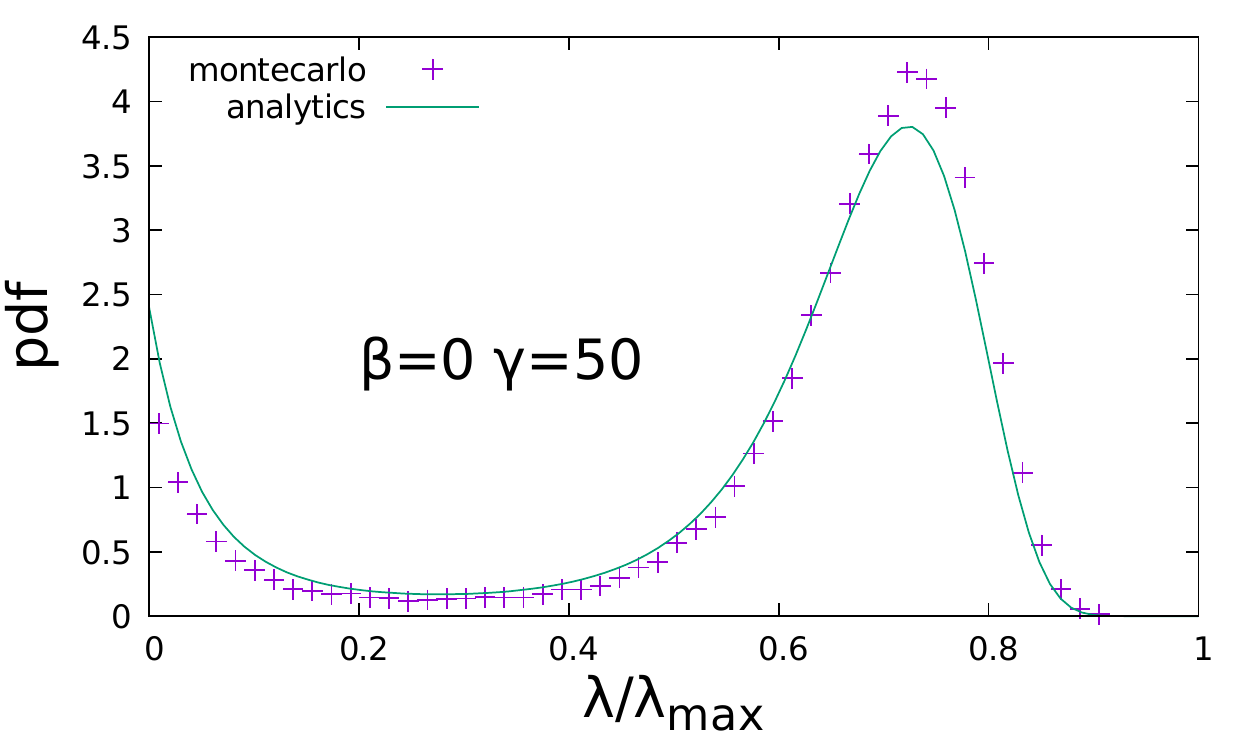}
\caption{Growth rate distributions for slow growing ($\beta=\gamma=0$), fast growing ($\beta=50, \gamma=0$) and coexisting states ($\beta=0, \gamma=50$ ) from analytical calculations (Legendre transform) and Monte Carlo computations on the  network under exam (see materials and methods)}
\label{fig:fig1}
\end{figure}
More in general, to each point of the phase diagram it corresponds a given distribution of metabolic fluxes, and in next section a simple analysis will be performed for a point inside the coexistence region for the model under exam, that is the central carbon metabolism of E Coli in glucose limited aerobic conditions and low dilution.

\subsection{Some examples of flux distributions at coexistence}

We consider the point ($\beta=0, \gamma=50$ ) in the coexistence region of the phase diagram. In Fig 3 we show scatter plots of a) the rate of oxygen consumption, b) carbon dioxide production and c) ATP synthase flux, all relative to the glucose uptake and scattered with the respect to the growth rate (relative to the maximum achievable one).
In qualitative agreement with stylizied facts, the slow growing phenotype results metabolically active as indicated by comparable level of ATP synthase flux
with respect to the fast growing phenotype (and increased variance). Further, the level of oxygen consumption and carbon dioxide production of the slow growing phenotype is higher with respect to the fast growing phenotype, clearly indicating this as a diversion of the carbon source intake  from the biomass.
However the absolute rate of glucose uptake of  fast and slow growing phenotypes are comparable within this simple framework, at odds with experiments with induced persistence where slower cells intake less carbon units.
\newpage 
\begin{figure}[h]
\centering
\includegraphics*[width=.45\textwidth,angle=0]{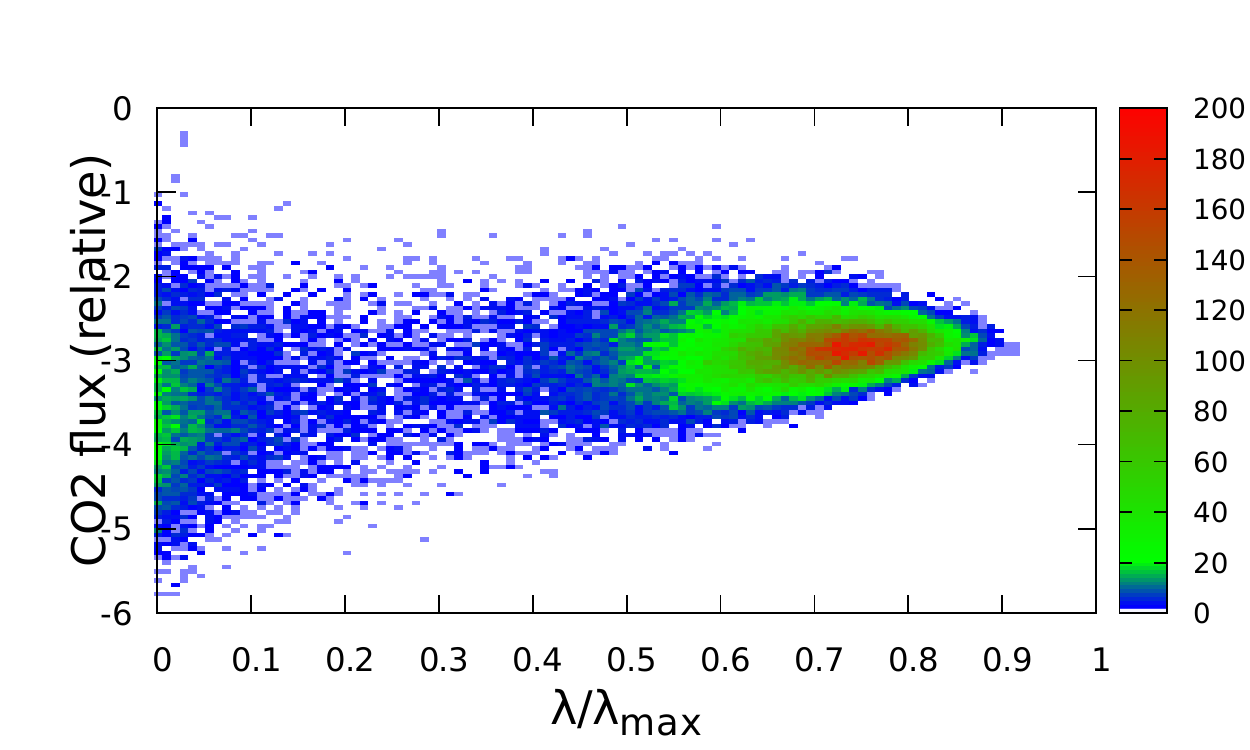}
\includegraphics*[width=.45\textwidth,angle=0]{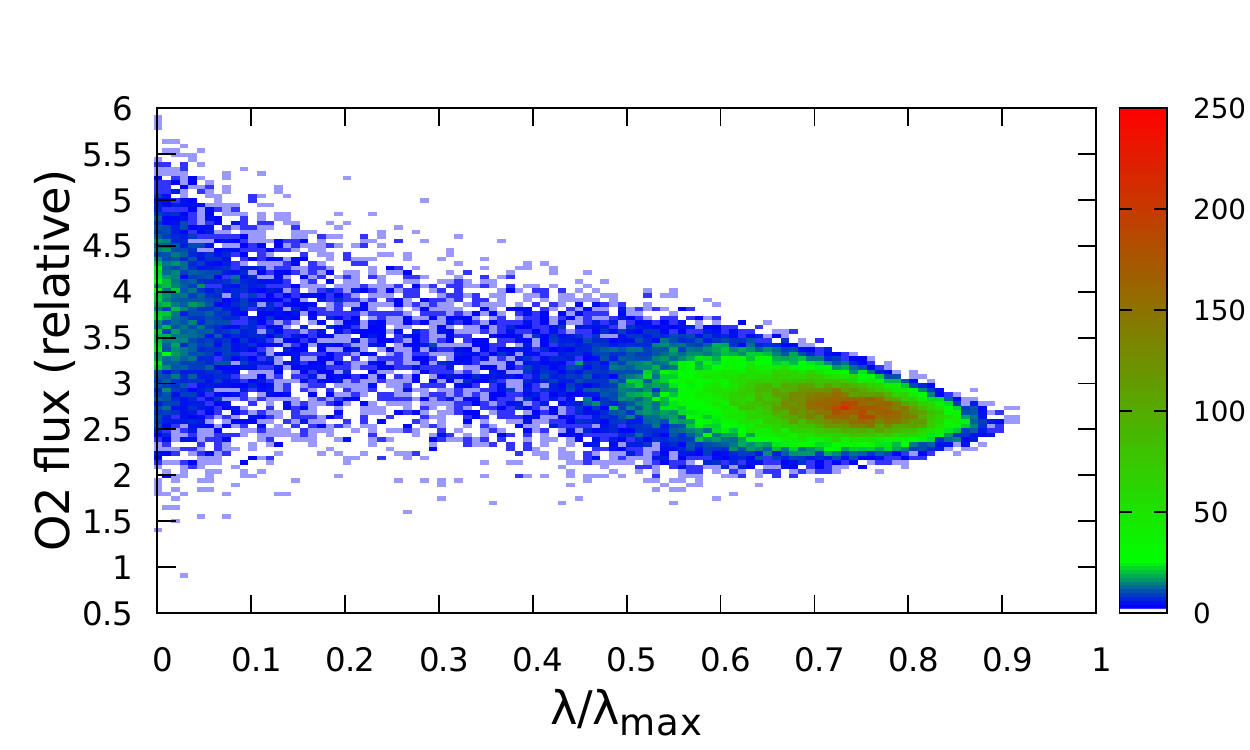}
\includegraphics*[width=.7\textwidth,angle=0]{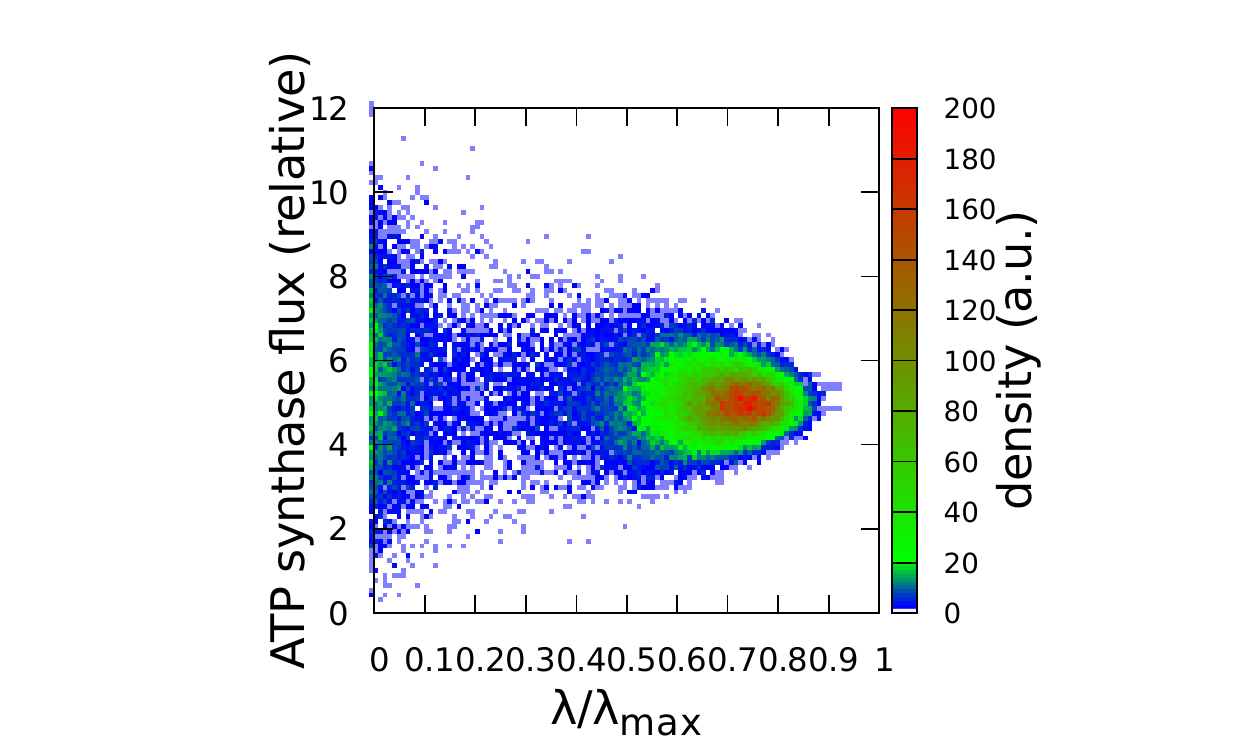}
\caption{Scatter plots of a) the rate of oxygen consumption, b) carbon dioxide production and c) ATP synthase flux, relative to the glucose uptake and scattered with the respect to the relative growth rate, for a point ($\gamma=50, \beta=0$) in the phase diagram in fig. 1. From numerical Monte Carlo calculations on the catabolic core of {\it E.Coli}.}
\label{fig:fig1}
\end{figure}

The flux distributions obtained at varying $(\beta,\gamma)$ can be then compared to experimental data in order to infer  coexistence of fast and slow growing phenotypes as we stress in the following section.

\subsection{Coexistence in standard conditions?}
Here it is examined the problem of inferring the parameters $(\beta^*,\gamma^*)$ that maximize the likelihood of data.  
For simplicty we consider a Gaussian approximation , i.e. 
if we have a subset of $K$ experimental fluxes, averages and variances, $\overline{\nu}_{i,exp},\sigma_{i,exp}$ we look at the $\chi^2$ as a function of the parameters $(\beta,\gamma)$ that is
\begin{equation}
\chi^2(\beta,\gamma) = \sum_i^K \frac{(\overline{\nu}_{i,exp} -\langle \nu_i \rangle_{\beta,\gamma} )^2}{\sigma^2_{i,\beta, \gamma}+\sigma^2_{i,exp}}
\end{equation}
A preliminary analysis has been performed to data for {\it E.Coli} cultures in standard  conditions (see materials and methods section). In Fig. 4 it is shown the $\chi^2(\beta,
\gamma)$ heat-map comparing data and predicted flux distributions in the plane $(\beta,\gamma)$. 
\begin{figure}[h]
\centering
\includegraphics*[width=0.9\textwidth,angle=0]{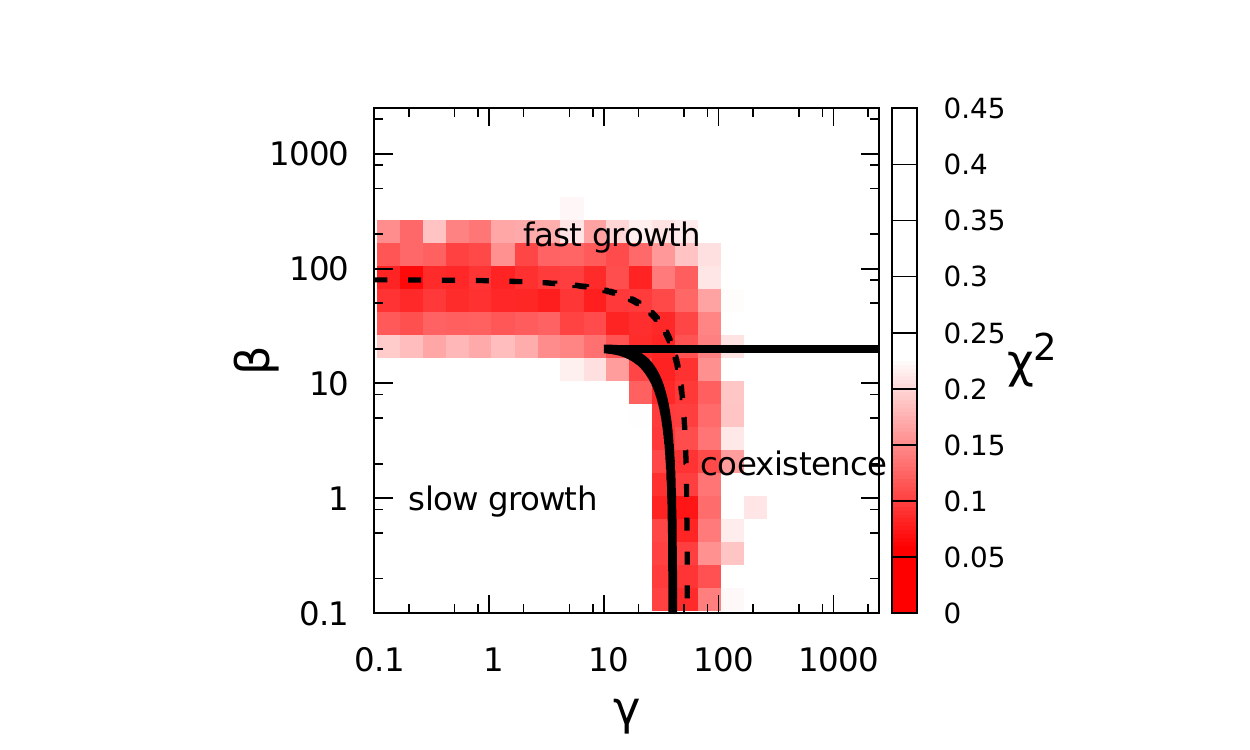}
\caption{$\chi^2(\gamma,\beta)$ heat-map comparing predicted flux distributions in the plane $(\gamma,\beta)$ and data referring to E Coli core metabolism for cultures grown in glucose limited aerobic conditions with growth/dilution rates below $0.4$h$^{-1}$, averages over $35$ technical replicates, control experiments retrieved from the database \cite{cecafdb}. Dashed line: $80 =\beta+1.5\gamma$.}
\label{fig:fig1}
\end{figure}
The inference correctly predict growing phenotypes, but  with a certain level of degeneracy for the inferred parameters, with low values of the $\chi^2$ that extend in the coexistence region. More precisely a line of $\chi^2$ minima develops approximately given by $80 =\beta+1.5\gamma$.

\section{Materials and methods}
The network analyzed is the catabolic core of a network reconstruction of {\it E. Coli} metabolism \cite{orth2010reconstruction},
that includes glycolysis, pentose phosphate pathway, Krebs cycle, oxidative phosphorylation and several nitrogen catabolic reactions.
The underlying polytope of steady states in glucose limited aerobic conditions turns out to be    a $D=23-$subdimensional variety in a space with $N=86$ (number of reactions) dimensions. States can be efficiently sampled with an hit-and-run Monte Carlo  Markov chain  upon handling of ill conditioning due to heterogeneous scales \cite{uniformell}, with rounding ellipsoids, for which a method due to L.Lovazs has been implemented \cite{lovazsbook}. The bias due to the Boltzmann-Gibbs factor enforcing the moments constraint has been implemented with a Metropolis-Hastings rejection rule for the proposed step during the hit-and-run dynamics \cite{hastings1970monte}. The flux data refer to E. Coli cultures grown in glucose limited minimal medium in aerobic conditions at low dilution/growth rates below $0.4$h$^{-1}$, i.e. the threshold for the acetate switch.
The sample consists of 35 technical replicates collected from control experiments retrieved in the database \cite{cecafdb} (same dataset analyzed in \cite{de2017statistical}).

\section{Conclusions}
In this work it has been shown that a simple phenomenological approach in which the first two moments of the growth rate in the space of steady states of a metabolic network are constrained can be used to model quantitatively bistability and coexistence of fast and growing cells in heterogeneous cell populations. This shows that bistability and coexistence can be an inherent feature of the central carbon core upon assuming control of the level of the average growth rate (optimization/repression) and its fluctuations (heterogeneity), providing in turn a quantitative map with metabolic flux distributions. It has been shown that this simple framework can account for stylized facts on persisters phenotypic heterogeneity, in particular their active metabolism at the level of ATP production, oxygen consumption rate and carbon dioxide production. 
This simple framework, however fails to predict lower level of carbon uptakes by the slow growing pehnotype as observed experimentally.
This last experimental fact refers however to stressed conditions, something that could be added phenomenologically by constraining the average covariance of biomass and uptakes, $\langle u \lambda \rangle$ (with a corresponding term in the  Boltzmann-Gibbs weight $\propto e^{\delta \lambda u}$). 
Such Maximum Etrnopy approach  can be further used to infer from flux data elusive coexistence and a preliminary analysis on data about E Coli cultures grown in glucose limited aerobic minimal medium  show degeneracy of parameters that extends in the coexistence region. This would require further investigations upon  complementing flux data with measures of single cell growth physiology, in particular rare events and in microfluidic devices where slower persisters are not taken over. Finally it is worth to mention that in general a Maximum Entropy approach shows  that constraining the first two moments of any flux in the metabolic space can lead  to a Van-Der-Waals picture with coexistence and bistability for the underlying flux, in turn providing an effective way to model single cell heterogeneity for any secondary metabolites production.

\section*{Acknowledgments}
The research leading to these results has received funding from the People Programme (Marie Curie Actions) of the European Union's Seventh Framework Programme ($FP7/2007-2013$) under REA grant agreement $n[291734]$. The Author thanks Prof. M. Heinemann for interesting discussions and kind hospitality in the University of Groningen.
\section*{References}    
\bibliographystyle{unsrt}
\bibliography{reference}

%\begin{thebibliography}{99}

%\end{thebibliography}

% Figure legends

\end{document}